\documentclass[twocolumn,showpacs,preprintnumbers,amsmath,amssymb]{revtex4}

\usepackage{graphicx,epsfig}
\usepackage{dcolumn}
\usepackage{bm}

\begin{document}
\topmargin .2cm
\title{ DILEPTON FROM QUARK-GLUON PLASMA AND QUARK-HADRON PHASE TRANSITION }
\author{S. SOMORENDRO SINGH , YOGESH KUMAR and D. S. GOSAIN}
\email{sssingh@physics.du.ac.in.}
\affiliation{ Department of Physics, University of Delhi, Delhi - 110007, INDIA}
\begin{abstract}
  A model of statistical quark-gluon plasma formation is considered.We look
the dilepton production at critical temperature $~T_{c}\sim170~Mev $ 
and completely free out temperature $T=150~MeV$ 
 with the 
initial temperature as $T_{0}=570,400 (250)~MeV$. Now we 
consider that quark mass
is  depending on
the coupling value through parameterisation factor 
of the fireball formation and temperature.
The rate of production 
is shown for 
invariant mass $M$  
at the particular value of $ E=2.0,3.0~ GeV$.It shows the 
significant production of leptons in this process for small value
of invariant mass. However, the quark-hadron phase transition is 
a very weakly changed in the entropy of the system during this 
process of hadronisation.
\end {abstract}
\maketitle
 The phase transition in high energy physics is predicted by the theory
of strong  
interactions so called  quantum chromodynamics(QCD)~\cite{wilczek}. 
This transition is observed at the temperature
 $ T_{c} \sim (150-170)~ MeV $ from the state of confined hadronic 
matter to the deconfined state of matter  so called 
quark-gluon plasma (QGP) in which quark and gluon 
become the relevant degrees of freedom of the plasma state . Presumably,
it is believed that it is really happened to be  
in the early universe of few micro seconds~$10\mu s$
just after the big bang and it seems to exist in the
interior core of neutron star too. Presently, the program of 
ultra-relativistic nucleus-nucleus collision indicates the presence 
of these strongly interacting matter at very high energy density and 
temperature . However, finding its evidence and ultimately proof
, for the creation and evolution of 
this deconfined state of matter 
is having a problem  to link the experimental observations
to the quantities measured in lattice calculations. But ,it has been 
recognized for a long time that electromagnetic radiation from
relativistic heavy ion collisions in these experiments would 
be a proper understanding for the formation of these hot and
dense plasma of quark and gluon , consequent to a quark-hadron
phase transition. But there are possible other probes to 
represent these new state of matter. Really they are enhanced
by the production of strangeness ~\cite{rafelski}, a suppression 
of $ J/\psi$~\cite{matsui} 
and 
radiation of photons and dileptons etc.~\cite{shuryak}. And
among these probable probes, dilepton and photon production are considered
to be most promising as they interact with electromagnetic force. So, they 
directly carry the whole informations of the plasma state without 
further scattering as their mean free path is very large enough to interact
and over a large  range of expected plasma temperature
, its production is observed throughout the evolution. 
Further, in the experimental status~\cite{NA50 collab}, it is reported 
that the $NA50$ experiments
 i.e $NA50$
collisions predict the excessive production of dileptons in the
intermediate mass range and $NA60$ collaborations~\cite{NA60 collab} 
in central $In-In$
collision at the CERN-SPS also show dimuon spectra from a hot
and dense hadron medium. In these way, several research groups have seen
these spectra for the same collision.
\par
 Thus, so far, the production of dilepton 
and photon in a QGP 
with a finite temperature is studied by many authors and 
these temperature is related to
 the energy density~$\epsilon$ given by the Stefan Boltzmamn~
$\epsilon\sim  T^{4}$ and their thermodynamic properties
are governed by $ T\frac{dP}{dT} - P = \epsilon ; P = \frac{1}{3}\sigma 
T^{4} - a T $ , where $ a $ is constant in non-perturbative effect 
in calculating the pressure. 
On these basis,  we study the dilepton production in a expanding
free baryonic QGP
from a statistical model of quark gluon plasma droplet
formation. Here the vanishing quark mass is depending on coupling value 
through 
some 
parameterisation factor which is again function of the temperature
 and it is again function of the temperature. 
Now, we obtained this parameterisation factor from the 
hydrodynamical fireball formation. It is really to control 
the dynamics of the plasma formation. 
It is obtained viz.~\cite{ramanathan}

\begin{equation}
\gamma_{q,g}=\sqrt{2}\sqrt{\frac{1}{\gamma_{g}^{2}}+\frac{1}{\gamma_{q}^{2}}}\, ,
\end{equation}
where, $\gamma_{g}=6~ or 8~\gamma_{q}$ and $\gamma_{q}=1/6$.This dynamical 
control value is very important to find the effective quark mass from 
its vanishing value . It is calculated by the technique 
of Braaten and Pisarski~\cite{pisarski} to 
remove the massless quark to be some factor of
this thermal temperature, depending on these parameterisation.Thus,
we obtain the production rate of the dileptons with low invariant mass
and the dilepton yeild with the evolution time $\tau$. 
Secondly, we look the quark-hadron phase transition from the free 
energy of these hydrodynamical model of QGP formation with this effective
quark mass. To see this transition, we 
calculated the thermodynamic properties of these system
and we compare these results with the transitional nature of the vanishing 
quark mass and the thermal dependent quark mass. After all, we show
the conclusion of these results with the standard results produced by 
many authors. 
\par
{\bf Dilepton production from QGP}: There are  good numbers of
research works in dilepton production from quark-gluon plasma 
and hadronic system. 
Whenever there is a new model for plasma evolution, then its impact on
dilepton and virtual photon production are assessed.
Moreover, it is expected ,
that in heavy-ion collisions such as RHIC at BNL and SPS at CERN,~\cite{nayak} 
the thermal production of dileptons will be more than
other process.  Thus ,
on these basis of these information, we focus our
model of dilepton production from the thermalised
quark-gluon plasma having initial temperature~ $ T_{0}=570 ,~400~(250) MeV$
to the transition temperature~ $T_{c}=(150-170) ~MeV$ , before 
it completely
freeze out to hadrons. So, we use the dominant
reaction for thermal emission of dilepton pairs~\cite{ruuskanen} 
is the Drell-Yan
mechanism $q\bar{q}\rightarrow l^{+}l^{-}$ as 
annihilation process or
$q(\bar{q})g \rightarrow q(\bar{q})+ l^{+}l^{-}$ as compton process. 
But, we 
exclusively use  $q\bar{q}\rightarrow l^{+}l^{-}$  reaction for 
this calculation. 
The dilepton production rate $\frac{dR}{d^{4}X}$(i.e the number of 
dilepton produced per space time volume  is given by:
\begin{eqnarray}
\frac{dR}{d^{4}X}&=&\int \frac{d^{3}p_{1}}{(2 \pi)^{3}} 
\frac{d^{3}p_{2}}{(2 \pi)^{3}} f_{q}(p_{1},T) 
f_{\bar{q}}(p_{2},T)\nonumber 
\\
&\times &v_{q\bar{q}} \sigma_{q\bar{q}}(M^{2}) 
\end{eqnarray}
where, $f_{q}(p_{1},T)$
and $f_{\bar{q}}(p_{2},T)$ are distribution functions of quark and antiquark.
$v_{q\bar{q}}$ is the relative velocity between quark and 
antiquark pair. $P_{\mu}$ is the four momentum of the lepton
pair,~$ E~(=\sqrt {M^{2}+P_{T}^{2}})$~is the energy,
$P_{T}$ is transverse momentum and~$M^{2}=P^{\mu}P_{\mu}$ 
is invariant lepton pair mass.$\sigma_{{q\bar{q}}
{\rightarrow}{l\bar{l}}}$ is the electromagnetic 
annihilation cross section. It is given as:
\begin{eqnarray}
\sigma_{q\bar{q}\rightarrow l^{-}l^{+}}(M^{2})
&=&\frac{4\pi \alpha^{2}}{3} \sum_{f=1}^{2}
(\frac{e_{f}}{e})^{2} \frac{1}{M^{2}} \nonumber \\ 
&\times&(1-\frac{4 m^{2}}{M^{2}})^{-1/2} 
\sqrt{(1-\frac{4 m_{l}^{2}}{M^{2}})} \nonumber \\
&\times &(1+2\frac{m^{2}+m_{l}^{2}}{M^{2}}+4\frac{m^{2}m_{l}^{2}}{M^{2}})
\end{eqnarray}
where $(\frac{e_{f}}{e})^{2} =\frac{5}{9}$, the electric charge of the quark 
in the units of the electron charge, $\alpha=\frac{1}{137}$ and $m_{l}$ is the 
lepton mass but we consider the lepton mass is zero(ie $m=0$). 
Substituting  the frequency distribution functions for quark
and antiquark in the  equation  $(2)$ , we integrate 
over $q$ and $\bar{q}$ momentum and subsequent
integrating over space and time with
the change in variable $d^{4}x=dx_{p}\tau d\tau dy$
, we obtain the dilepton production rate as:
\begin{eqnarray}
 \frac{dR}{dM^{2}dy}& =& \frac{5 \alpha^{2}}{6 \pi^{2}}
\frac{\tau^{2}R^{2}T_{0}^{6}}{M^{4}}
 (1+\frac{2 m^{2}}{M^{2}})\nonumber \\
&\times &\int_{\frac{M}{T_{0}}}^{\frac{M}{T_{c}}} z^{4}K_{1}(z)dz  
\end{eqnarray}
where $K_{1}(z)$ is bessel's function. Its integral can be obtained as
$ -z^{2}(8+z^{2})K_{0}(z)-4z(4+z^{2})K_{1}(z)$ and $z=M/T$. Here, $R$ 
is size of the quark droplet. 
  The creation and evolution 
of the plasma is very short time i.e around $\tau\sim 1 fm/c$ ~ and ,
 finally transits upto temperature $T_{c}\sim (150-170) MeV$.
In the above equation, the quark mass is 
dependent on the temperature 
which is given by $ m^{2}(T)= \frac{1}{6} g^{2}T^{2}$, where $'g'$ is 
coupling parameter and it is obtained as~$[7]$:
\begin{equation}
    g^{2}=\frac{16\pi}{27}
         \frac{1}
          {
           ln(1+\frac{p^{2}}{\Lambda^{2}})
          }
\end{equation}
with the QCD parameter $\Lambda = 150 MeV $ and 
$ p=(\frac{\gamma_{q,g}N^{\frac{1}{3}}T^{2}\Lambda^{2}}{2})^{\frac{1}{4}}$
is low momentum cut-off value with $N=\frac{16\pi}{27}$. It is found that 
$g$ is a slighly strong compared to the other coupling value. 
\begin{figure}
\resizebox*{3.1in}{3.1in}{\rotatebox{360}{\includegraphics{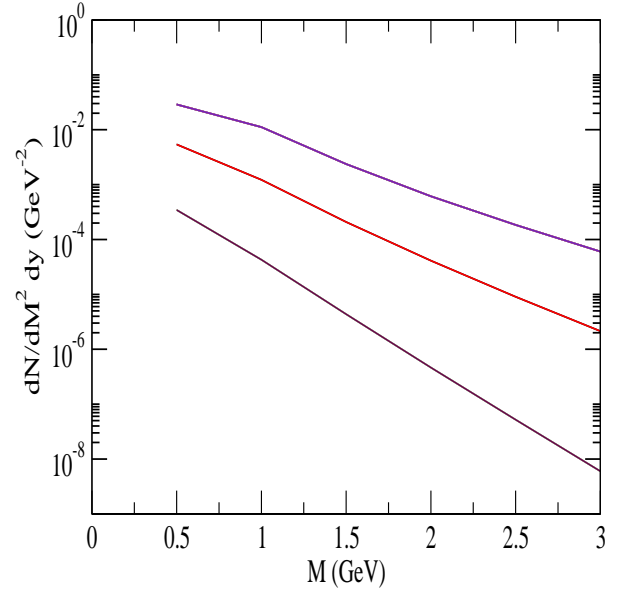}}}
\vspace*{0.5cm}
\caption[]{
The dilepton emission rate at~
$ E=3.0 GeV,$ ~ with initial temperature
$ T_{0}=0.57~GeV,~0.4~GeV$ and~$ 0.25~GeV~ $ at~$ T_{c}=0.15 GeV$. At$ E=2.0~GeV$ and 
$E=3.0~GeV$ the plot is same with the same initial temperature.
}
\label{scaling}
\end{figure}
\begin{figure}
\resizebox*{3.1in}{3.1in}{\rotatebox{360}{\includegraphics{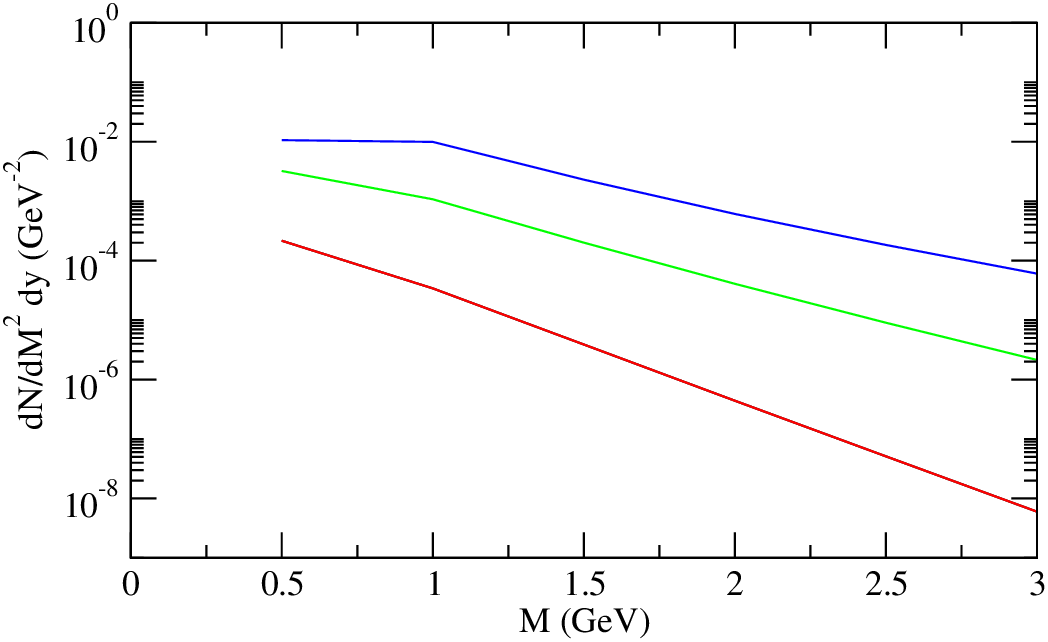}}}
\vspace*{0.5cm}
\caption[]{
The dilepton emission rate at~
$ E=3.0 GeV,$ ~ with initial temperature
$ T_{0}=0.57~GeV,~0.4~GeV$ and~$ 0.25~GeV~ $ at~$ T_{c}=0.17 GeV$. At $ E=2.0~GeV
~$and~$ E=3.0~GeV$  the plot is the same with the same initial temperature.
}
\label{scaling}
\end{figure}

\begin{figure}
\resizebox*{3.1in}{3.1in}{\rotatebox{360}{\includegraphics{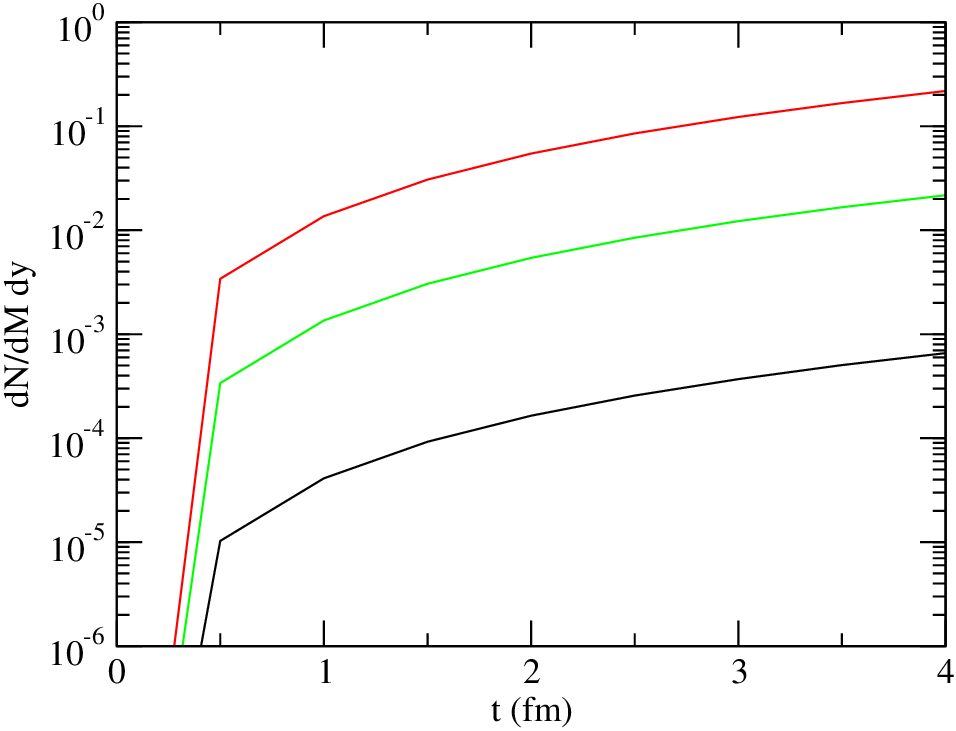}}}
\vspace*{0.5cm}
\caption[]{
The integrated yield as function of t 
at$ E=3.0 GeV,$ ~ with initial temperature
$ T_{0}=0.57~GeV,0.4~GeV$ and~$ 0.25~GeV~ $ at~$ T_{c}=0.15 GeV$. At $ E=2.0~GeV
~$and~$ E=3.0~GeV$  the plot is the same with the same initial temperature.
}
\label{scaling}
\end{figure}
\begin{figure}
\resizebox*{3.1in}{3.1in}{\rotatebox{360}{\includegraphics{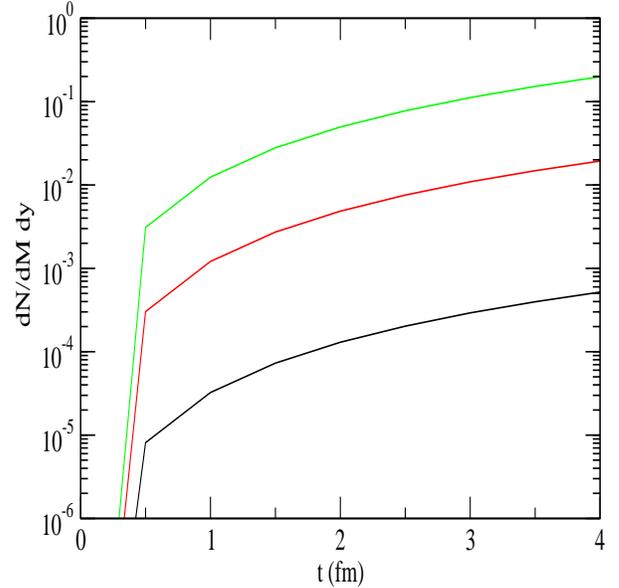}}}
\vspace*{0.5cm}
\caption[]{
The integrated yield as function of t at~
$ E=3.0 GeV,$ ~ with initial temperature
$ T_{0}=0.57~GeV,0.4~GeV$ and~$ 0.25~GeV~ $ at~$ T_{c}=0.17 GeV$. At $ E=2.0~GeV
~$and~$ E=3.0~GeV$  the plot is the same with the same initial temperature.
}
\label{scaling}
\end{figure}
\begin{figure}
\resizebox*{3.1in}{3.1in}{\rotatebox{360}{\includegraphics{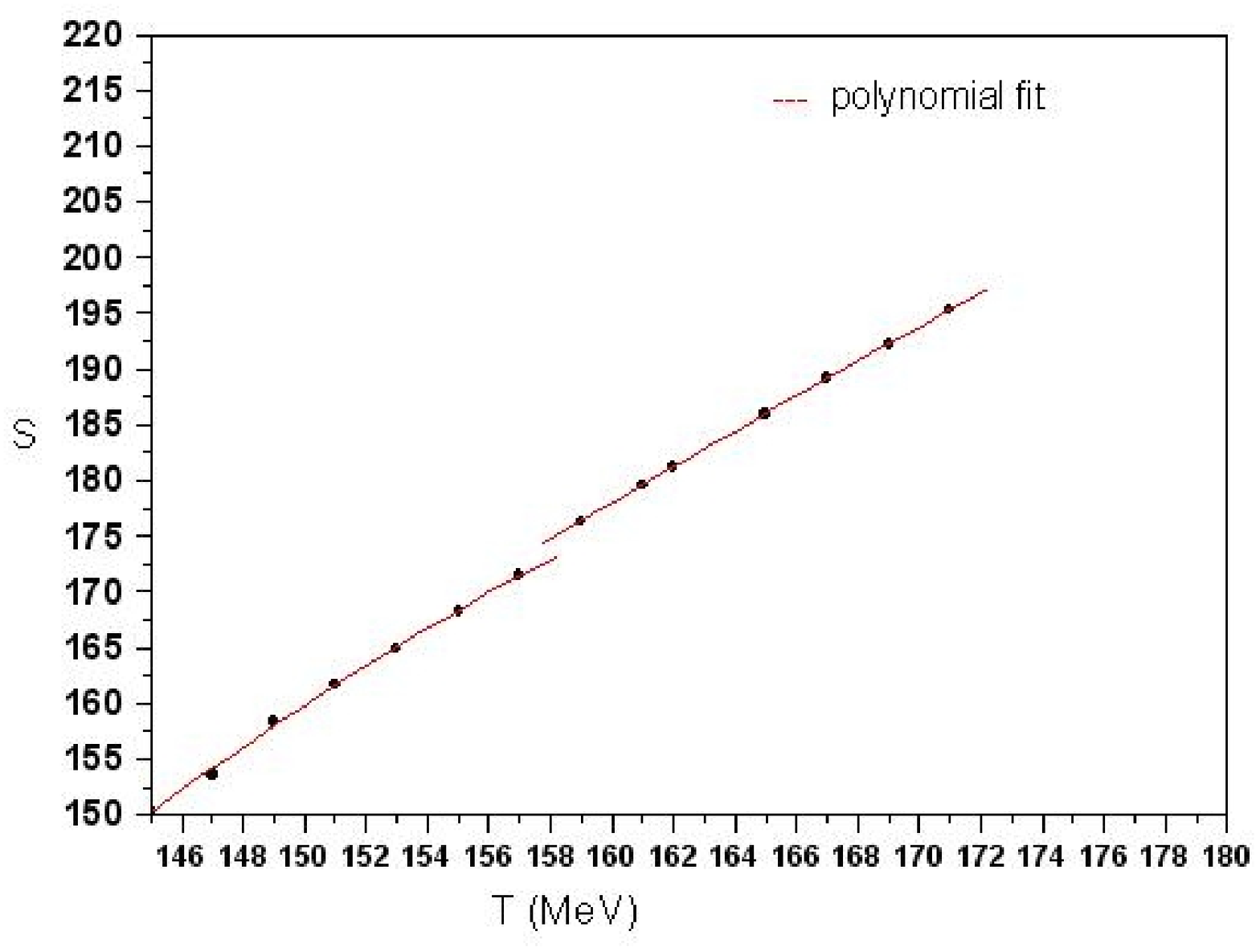}}}
\vspace*{0.5cm}
\caption[]{Variation of~ $S$~with temperature ~$T$~ 
at $\gamma_{g}=8 \gamma_{q},~\gamma_{q}=1/6.~$ }
\label{scaling}
\end{figure}
\begin{figure}
\resizebox*{3.1in}{3.1in}{\rotatebox{360}{\includegraphics{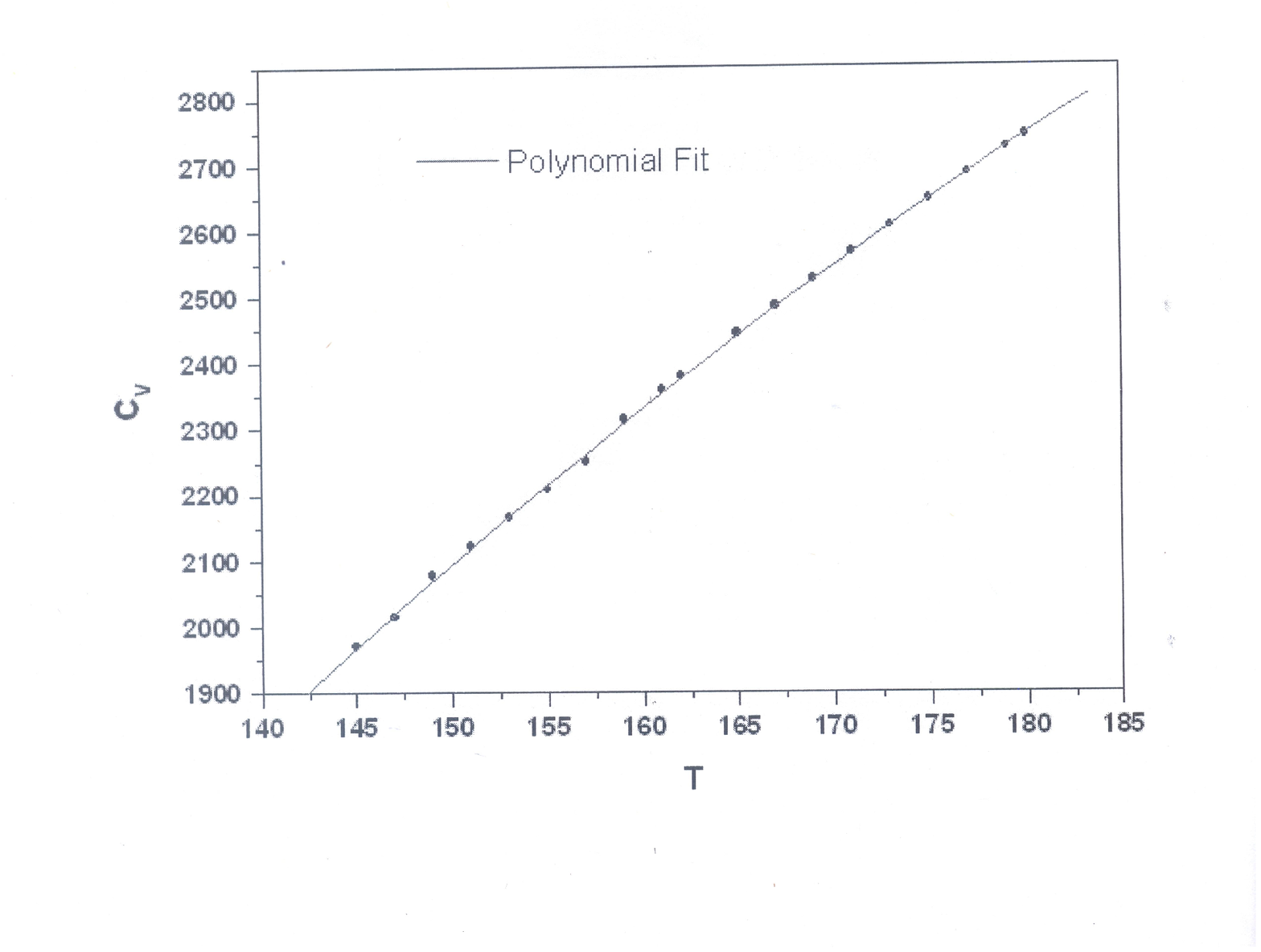}}}
\vspace*{0.5cm}
\caption[]{ variation of specific heat $C_{v}$ with temperature $T$
at $\gamma_{g}=8 \gamma_{q}, \gamma_{q}=1/6.$
}
\label{scaling}
\end{figure}
\begin{figure}
\resizebox*{3.1in}{3.1in}{\rotatebox{360}{\includegraphics{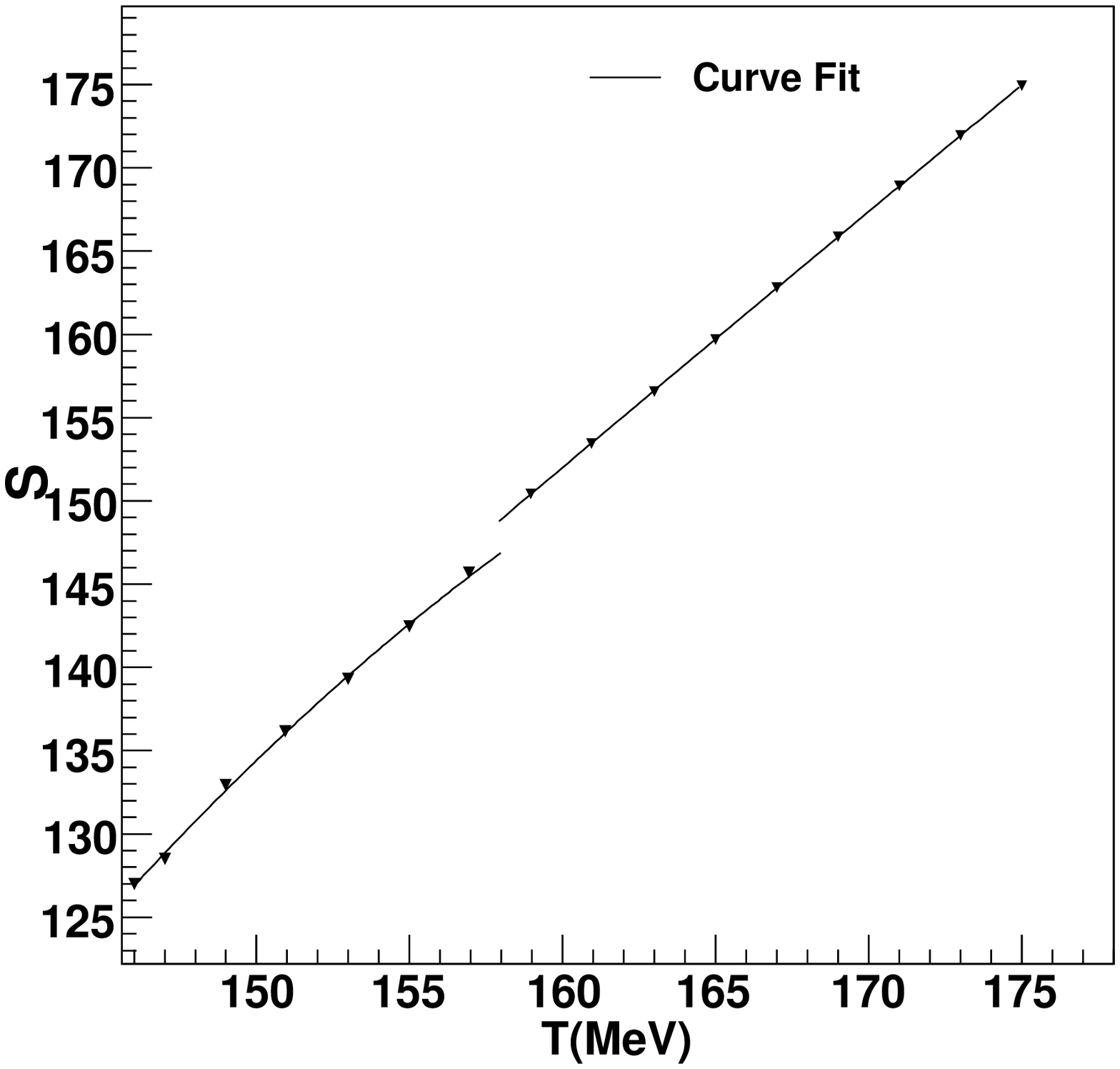}}}
\vspace*{0.5cm}
\caption[]{Variation of~ $S$~with temperature ~$T$~
at $\gamma_{g}=8 \gamma_{q},~\gamma_{q}=1/6~$~ with thermal mass.}
\label{scaling}
\end{figure}
\begin{figure}
\resizebox*{3.1in}{3.1in}{\rotatebox{360}{\includegraphics{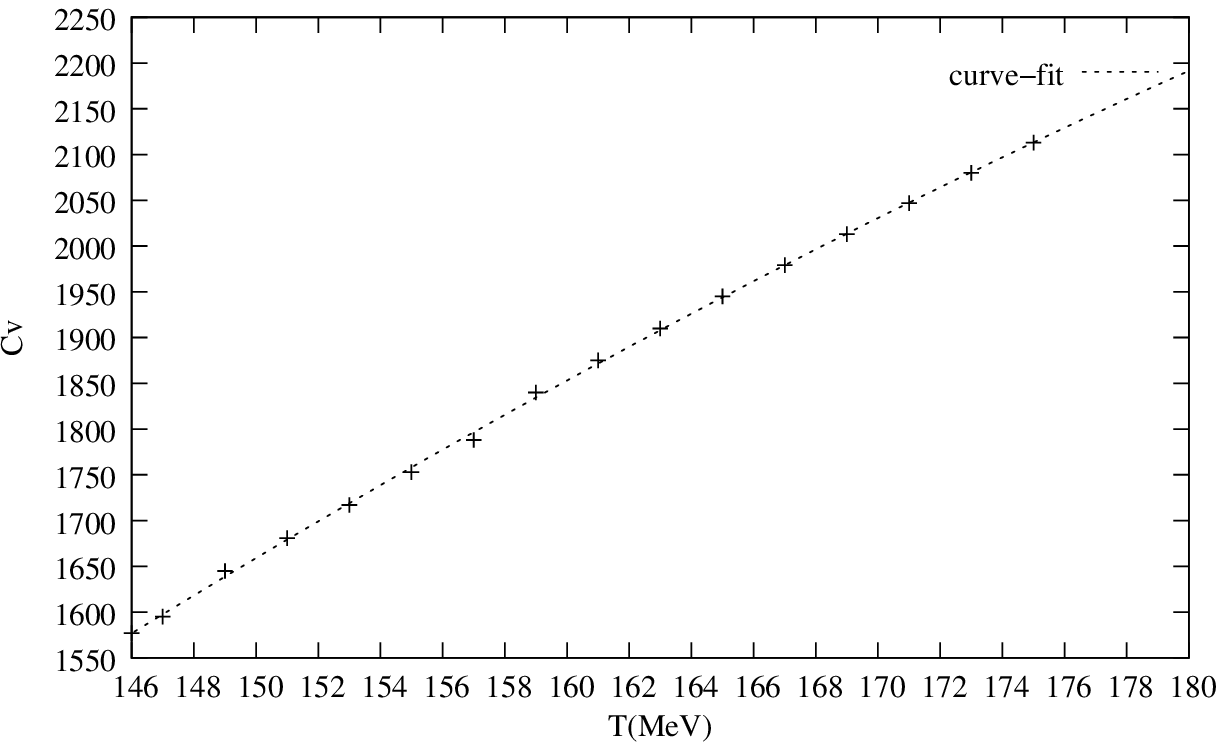}}}
\vspace*{0.5cm}
\caption[]{ variation of specific heat $C_{v}$ with temperature $T$
at $\gamma_{g}=8 \gamma_{q}, \gamma_{q}=1/6$~ with thermal mass.
}
\label{scaling}
\end{figure}
\par
{\bf Quark-Hadron Phase Transition}: The phase transition from 
quark-gluon
to hadronic phase is evaluated through the free energy of system of 
fermions (upper sign) or boson (lower sign) at 
the temperature $T$. The grand canonical free energy of these system 
is given by the relation:
\begin{equation}
       F_{i}=\mp T g_{i}\int dp \rho_{i}(p)ln(1\pm \exp(-\sqrt{m_{i}^{2}+p^{2}}/T))
\end{equation}
where $\rho_{i}(p)$ is density of states of the particular particle,$i$ (quark
,gluons) with momentum between $p$ and $p+dp$ in a spherically symmetric
situation, and $g_{i}$ is the degeneracy factor (colour and spin degeneracy)
which is $6$ for quark and $8$ for gluons. In a similar manner, the free energy for the pionic environtment is given as:
\begin{equation}
       F_{\pi}=(3T/2\pi^{2}) v \int_{0}^{\infty} p^{2}dpln(1- \exp(-\sqrt{m_{\pi}^{2}+p^{2}}/T))
\end{equation}
and for interfacial energy, it is
\begin{equation}
     F_{interface}=\frac{1}{4}R^{2}T^{3}\gamma_{q,g}
\end{equation}
From these above free energies, we can obtain the standard thermodynamic
properties of the system. They are as follows
\[ Entropy, S=-(\frac{\delta F}{\delta T})\], 
and \[ Specific~ heat, C_{v}= T (\frac{\delta S}{\delta T})_{v}
\]
Now , we compare the transitional nature of these system
with the vanishing quark mass
and the thermal dependent quark mass.
\par
{\bf Results and Conclusions}: In this short article , we attempt to 
evaluate the dilepton production from a statistical model
of QGP fireball formation and its quark-hadron phase transition. Now, we 
consider the vanishing quark mass is dependent on the thermal 
temperature and it is replaced by this thermal mass model obtained
through the parameterisation factor of fireball formation .
First we discuss the results of the evaluated equation~$(4)$.
They are shown 
in the Figs.$[1-4]~ $~with their 
corresponding variations with production rates. 
The production rate of the dilepton at $E=2.0 ~GeV$ and $E=3.0~GeV$
with variation of dilepton invariant mass
is shown in the Fig.$[1]$ and Fig.$[2]$.There are
emission of these dilepton particles at 
low invariant mass region viz. ~$M=(0-3.0)~GeV$. 
The result is quite in agreement with many results for the low 
invariant mass produced by other authors.In our fig.$[1]$, the production 
of dilepton is shown 
with initial temperature $T_{0}=570,400$ and $250~MeV$ at the 
free-out temperature $T=150 ~MeV$. Again in fig.$[2]$ ,it is the
spectra at transition temperature $T_{c}= 170 ~MeV$.
The comparation of production rate in these two figures
are slightly different.
The production rate at free out is 
a bit high compared with the transition $T_{c}=170~MeV$. 
It means ,~near the hadronic
phase its production is more enhanced. But
the production is indistinguishable for the different 
values of the energies at the same initial temperature $T_{0}$~and 
transition temperature $T_{c}$. It means, the rate 
is almost same with these 
two energies $E=2.0, 3.0~ GeV$. Again, we look the integrated yeild with 
variation of time from $\tau=0~ fm~ $to $\tau=3.0~fm$ for
these transition and free out temperature. There is a very slight
different for this integrated yeilds too. But it shows very nice 
scenarios with these time region. This means there is overall 
agreement with those results of Zejun He et al.\cite{zejun he}.    
Moreover, for the transition it is shown in figs.$[5-8]$.
Again, it is reported that the process
of transition is very weakly first order
around the transition temperature ~$T_{c}=(150-170)~MeV$ with
vanishing quark mass 
in figs. ~$[5-6]$~ and transition with the 
temperature dependent quark mass is clearly shown in figs.~$[7-8]$.
  From this, we look
what it 
is happening to the order of transition if the quark mass is replaced 
by the thermal
dependent mass . Now  it is found that the sitiution is same with 
the transition in entropy of the massless quark with a very
clear jump around the temperature~$ T_{c}\sim(160\pm 5)MeV$ with
less amplitude of the entropy.
 So, we could justify as the first order transition in 
this model of quark mass dependent on temperature and 
parameterisation factor.
   
\acknowledgements
  We are very thankful to Dr. R. Ramanathan and Dr. K. K. Gupta 
for their constructive 
suggestions and discussions for whole of this work. One of the author
, Yogesh kumar is research student of this department.He would like to 
express 
his gratitude to Rajiv Gandhi
fellowship for the financial support.

\end{document}